\documentclass[showpacs,aps,pra,floatfix,article,superscriptaddress]{revtex4-1}
\usepackage[english]{babel}
\usepackage[utf8]{inputenc}
\usepackage{amsmath,amssymb,amsfonts,physics}
\usepackage{ragged2e}
\usepackage{graphicx}
\usepackage [T1] {fontenc}
\usepackage{bbold}
\usepackage{cases}
\usepackage[all]{xy}
\usepackage{hyperref}
\usepackage{xcolor}

\newcommand{\orcid}[1]{\unskip}

\usepackage{hyperref}
\usepackage{amsfonts, vmargin}
\usepackage[us]{datetime}

\begin{document}
\title{Lax Connection and Conserved Quantities of Quadratic Mean Field Games}

\author{Thibault Bonnemain\orcid{0000-0003-0969-2413}}
\affiliation{Université Paris-Saclay, CNRS, LPTMS, 91405, Orsay, France}   
\affiliation{LPTM, CNRS, Universit\'e Cergy-Pontoise, 95302 Cergy-Pontoise, France}
\affiliation{Department of Mathematics, Physics and Electrical Engineering, Northumbria University, Newcastle upon Tyne, United Kingdom}

\author{Thierry Gobron\orcid{0000-0001-6641-671X}}
\affiliation{LPTM, CNRS, Universit\'e Cergy-Pontoise, 95302 Cergy-Pontoise, France}
\affiliation{CNRS UMR 8424,  Universit\'e de Lille, Laboratoire Paul Painlev\'e,  59655 Villeneuve d'Ascq, France.}

\author{Denis Ullmo\orcid{0000-0003-1488-0953}}
\affiliation{Université Paris-Saclay, CNRS, LPTMS, 91405, Orsay, France}  

\begin{abstract}
Mean Field Games is a new field  developed simultaneously in applied mathematics and engineering in order to deal with the dynamics of a large number of controlled agents or objects in interaction. For a large class of these models, there exists a deep relationship between the associated system of equations and the non-linear Schrödinger equation, which allows to get new insights on the structure of their solutions. In this work, we deal with related aspects of integrability for such systems, exhibiting in some cases a full hierarchy of conserved quantities, and bringing some new questions which arise in this specific context.

\vspace{2cm}

\end{abstract}

\maketitle

\section{Introduction}
Mean Field Game (MFG) is a rather recent theoretical framework specifically developed to address complex problems of game theory when the number of players becomes large \cite{LasryLions2006-1,LasryLions2006-2,LasryLions2007,Huang2006}. Accordingly, they have natural applications in various fields, ranging from
finance \cite{Lachapelle2014,Cardaliaguet2017,Carmona-ctrl2013} to sociology \cite{Achdou2016,Achdou2014,GueantLasryLions2010} and
engineering science \cite{kizilkale2016collective,kizilkale2019integral,meriaux2012mean}, and more generally whenever optimization issues involve a large number of coupled subsystems.

Such games can be characterized by the coupling between a time-forward
diffusion process, for a density $m(\vec x,t)$ of agents with state variables $\vec x \in
\mathbb{R}^n$ at time $t$,  and an optimization process
resulting in a value function $u(\vec x,t)$ constructed backwards in time. 
In the simplified case of quadratic mean field games
(see \cite{ULLMO20191} for a suitable introduction for physicists) this 
construction leads to a system of two coupled equations: a forward Fokker-Planck equation for the density and a backward Hamilton-Jacobi-Bellman equation for the value function
\begin{equation}\label{MFGeq}
\left\{
\begin{aligned}
&\partial_t m - \frac{1}{\mu}\vec \nabla.\left[m\vec \nabla u\right]-\frac{\sigma^2}{2}\Delta m=0 \\
&m(\vec x,t=0)=m_0(\vec x)\\
&\partial_t u + \frac{\sigma^2}{2}\Delta u - \frac{1}{2\mu}||\vec \nabla u||^2=V[m] \\
&u(\vec x,t=T)=c_T[m](\vec x)
\end{aligned} \right. \; ,
\end{equation}
where $\sigma$ and $\mu$ are positive constants. In  such a case, the two PDE's are coupled through two terms: in the first equation, the optimization process enters through the drift velocity  for the density, ${\vec a}(\vec x,t) $, which optimal value is proportional to the gradient
of the value function as ${\vec a}(\vec x,t) = - \vec\nabla
u(\vec x,t)/\mu$; in the second one, the source term for the value function in the right hand side derives from mean field type interactions and involves a functional of the density $m$ at time $t$, $V[m(\cdot,t)](\vec x)$ (that may
also have an explicit dependence in $\vec x$). The atypical forward-backward structure, with mixed initial and final boundary conditions, leads to new
challenges when trying to characterize solutions, either analytically or
numerically.

This paper is dedicated to the study of a class of so-called \textit{integrable quadratic Mean Field Games} which, in principle, can be solved entirely analytically. The main motivation behind this work is to make use of the deep connection between quadratic MFG and the non-linear Schrödinger (NLS) equation - which is integrable under some conditions - in order to grasp new formal results for this forward-backward system of equations \eqref{MFGeq}. These integrable games are very specific  but can be seen as limiting regimes of more general problems, such as the ones considered in \cite{bonnemain2019schrodinger}. Very few realistic situations can accurately be described by such games, but their interest lies in the fact that they can serve as reference models for  more general approaches.

By \textit{integrable quadratic Mean Field Games} we specifically refer to games described by the system of MFG equation in 1+1 dimensions with $V[m] =g m$, already studied in \cite{bonnemain2019universal} in the repulsive case ($g<0$), featuring linear local interactions but no explicit dependence on the position and in particular no external potential:
\begin{equation}\label{iqMFG}
\left\{
\begin{aligned}
&\partial_t m - \frac{1}{\mu} \partial_x \left[m\partial_x u\right]-\frac{\sigma^{2}}{2}\partial_{xx} m=0 \\
&m(x,t=0)=m_0(x)\\
&\partial_t u + \frac{\sigma^2}{2}\partial_{xx} u - \frac{1}{2\mu} (\partial_x u)^2=g\,m \\
&u(x,t=T)=c_T[m](x) 
\end{aligned} \right. \; .
\end{equation}
This can be seen as a particular, admittedly very simple, case of the population dynamics model introduced by O. Guéant in 2010 \cite{GueantLasryLions2010}, in which players have no preferences whatsoever for a given state $x$ (representing maybe a physical position, capital, beliefs etc.), but only care about the number of other players in their close vicinity. The sign of the constant $g$ monitors the type of interactions between players. A positive sign would correspond to attractive interactions (herding effect, peer pressure, etc.) while a negative value would describe repulsive interactions (collective exploration, anti-conformism, etc.). Both instances has been the subject of extensive discussions (respectively \cite{igorthese,Swiecicki-prl-2016,ULLMO20191} and  \cite{bonnemain2019universal,bonnemain2019schrodinger}) but  the present considerations on integrability are new.

The aim of this paper is to show that these games are (completely) integrable in the Liouville sense \cite{liouville1855note}. They can be seen as infinite dimensional Hamiltonian systems, for which  an infinite number of commuting Poisson invariants can be constructed. These conserved quantities are in \textit{involution} and are known as \textit{first integrals of motion}. Another, more geometrical, way of saying this is that there exists a regular 
foliation of the phase space by invariant manifolds, such that the Hamiltonian vector fields associated with the invariants of the foliation span the tangent  space. By the Liouville-Arnold theorem \cite{arnol2013mathematical}, for such systems there exists a canonical transformation to \textit{action-angle variables} (as in preserving Hamilton's equations). In this system of coordinates, the Hamiltonian  depends only on the action variables (which are equivalent to the first integrals of motion), while the dynamics of angle variables is linear. If this canonical transform is explicitly known, the system can be solved by quadratures, in which case these games can be considered as ``completely solvable analytically''. 

In this paper, we will prove the existence of the integrals of motion. This constitutes only a first (and probably the simplest, albeit non-trivial) step in the aforedescribed procedure but also the more useful. Computing conserved quantities has natural implications outside of the realm of integrable systems and can serve, for instance, as a basis for variational approaches. Section \ref{sec:nlsflism} contains the basic informations: it introduces the non-linear Schrödinger representation and provides the reader with a physical interpretation of the first integrals of motion as well as a general recipe, without justification, on how to compute them. The aim of this section is to provide readers  with straightforward and immediately applicable results, without diving too much  into formal issues. In section \ref{sec:intflism}, we discuss essential notions of integrable systems, e.g. the zero-curvature representation, necessary to derive explicit expressions for the integrals of motion. Section \ref{sec:1stint} provides a direct computation of conserved quantities and a proof of the fact that they are all in involution, using a generalisation of the Hamiltonian formalism to infinite dimensional systems. We conclude this paper with a summary of the results and a discussion on the next steps required to completely solve the problem, namely the computation of action-angle variables.

\section{Schrödinger representation of quadratic Mean Field Games}
\label{sec:nlsflism}

\subsection{Canonical change of variables}

The integrability of quadratic MFG can be traced back to the Schrödinger representation \cite{Gueant2012,ULLMO20191} of Eqs.~\eqref{iqMFG} 
\begin{equation}\label{intNLS}
\left\{
\begin{aligned}
&- \mu\sigma^2\partial_t\Phi= \frac{\mu\sigma^4}{2}\partial_{xx}\Phi + gm\Phi \\
&+ \mu\sigma^2\partial_t\Gamma=\frac{\mu\sigma^4}{2}\partial_{xx}\Gamma + gm\Gamma
\end{aligned} \right. \; ,
\end{equation}
obtained by performing a Cole-Hopf like transform 
\begin{equation}
\left\{
\begin{aligned}
u(t,x) & = - \mu\sigma^2 \log \Phi(t,x)  \\ 
m(t,x) & = \Gamma(t,x) \Phi(t,x)  
\end{aligned} \right. \; .
\end{equation}
NLS being integrable in the absence of external potential, we may expect that its MFG counterpart, equations \eqref{intNLS}, has the same property.

One of the most powerful methods when it comes to exploiting the integrability of MFG equation was first introduced by V. Zakharov and A. Shabat in their seminal paper of 1972 \cite{shabat1972exact}. It presents what would later be known as the \textit{inverse scattering transform} (IST) and constitutes the basis of soliton theory \cite{novikov1984theory}. This paper represents a first step in adapting this method, in its modern formulation, to MFG and relies heavily on the book  by L. Faddeev and L. Takhtajan \cite{faddeev2007hamiltonian}. It presents the IST formalism and how it can be applied to MFG, but does not provide a solution to the system of equations \eqref{intNLS}. Instead it examines intermediate results, such as a way to generate first integrals of motion, and discusses some of the issues appearing in the context of MFG that will need to be addressed for further developments. 

\subsection{Action functional and Noether theorem}
\label{sec:noether}
One of the more immediate benefits of this alternative representation
is that it enables, in a fairly direct fashion, the introduction
of various methods and notions originally developed to study and
characterise problems of physics. Most notably, it brings forward the
concepts of action and energy to the context of MFG.
The system of Eqs.~\eqref{intNLS} can be obtained as stationarity conditions for
an action functionnal $S$ defined as
\begin{equation}\label{S}
S[\Gamma,\Phi]\equiv
\int_{0}^{T}dt\int_{\mathbb{R}}dx\left[\frac{\mu\sigma^2}{2}(\Gamma\partial_t\Phi-\Phi\partial_t\Gamma)-\frac{\mu\sigma^4}{2}\partial_x\Gamma\partial_x\Phi+\frac{g}{2}(\Gamma\Phi)^2\right] \; , 
\end{equation}
so that
\begin{equation}
\mbox{ Eq.~\eqref{intNLS}} \quad \Leftrightarrow \quad
\left\{
\begin{aligned}
\frac{\delta S}{\delta\Phi} & = 0 \\
\frac{\delta S}{\delta\Gamma} & = 0 
\end{aligned}
\right. \; .
\end{equation}
Existence of the action $S$ already implies, through Noether theorem, that conserved quantities  are associated with the explicit symmetries of the problem, the most notable example being an Energy
\begin{equation}\label{energy}
\begin{aligned}
E&=\int_{\mathbb{R}}dx\left[-\frac{\mu\sigma^4}{2}\partial_x\Gamma\partial_x\Phi+\frac{g}{2}\left(\Gamma\Phi\right)^2\right]\\
&=\int_{\mathbb{R}}dx\left[\frac{\sigma^2}{2}\left(\partial_x m\partial_x u + m\frac{(\partial_x u)^2}{\mu\sigma^2}\right)+\frac{g}{2}m^2\right]
\end{aligned}\; ,
\end{equation}
 which derives from the  invariance of the action under a time translation:
\begin{equation}\label{timetrans}
m(x,t)\rightarrow m(x,t+t') \quad \quad u(x,t)\rightarrow u(x,t+t') \; .
\end{equation}
 Two other relevant conserved quantities with a clear physical meaning are the (normalized) number of players
\begin{equation}\label{Numberplayers}
N=\int_{\mathbb R} m dx=\int_{\mathbb R} \Phi\Gamma dx=1 \; ,
\end{equation}
corresponding to $S$-invariance through  a shift of the value function $u$ by a constant
\begin{equation}
u(x,t)\rightarrow u(x,t) + u' \; ,
\end{equation}
and the momentum
\begin{equation}\label{momentum}
P = \frac{1}{2}\int_\mathbb{R}( \Gamma\partial_x\Phi-\Phi\partial_x\Gamma) dx =-\frac{1}{2}\int_\mathbb{R}\left[\partial_x m + \frac{2m}{\mu\sigma^2}\partial_x u\right]dx \; ,
\end{equation}
associated with invariance under a space translation
\begin{equation}\label{spacetrans}
m(x,t)\rightarrow m(x+x',t) \quad \quad u(x,t)\rightarrow u(x+x',t) \; ,
\end{equation}
under which the action $S$ is also invariant. The other conserved quantities exhibit more complicated expressions and their signification is usually more abstract.  Relying on Noether theorem to find those is impractical, and we need a more systematic way of generating conserved quantities.

\subsection{Recursion relations for the first integrals of motion}

A technical discussion on the derivation of explicit expressions for the conserved quantities can be found in Section \ref{sec:1stint},  using tools introduced in Section \ref{sec:intflism}. Hereafter, we provide a simple prescription for their construction, without demonstration.

For the sake of simplicity, we consider Mean Field Games with  fields $\Phi$ and $\Gamma$ decreasing sufficiently fast at infinity
\begin{equation} \label{eq:PhiGammaToZero}
\left\{
    \begin{aligned}
    & \lim_{x\rightarrow {\pm}\infty}\Phi(x,t)  = 0 \\
    &\lim_{x\rightarrow {\pm}\infty}\Gamma(x,t)  = 0
    \end{aligned}\right. \quad \Rightarrow \quad
    \left\{
    \begin{aligned}
    & \lim_{x\rightarrow {\pm}\infty} u(x,t) = +\infty \\
    &\lim_{x\rightarrow {\pm}\infty} m(x,t) = 0
    \end{aligned} \right. \; .
\end{equation}

As we shall show later, under this assumption every first integral of motion, denoted $Q_n$, can be written in the form
\begin{equation} \label{eq:Qn}
    Q_n=\int_{\mathbb{R}}w_n\Phi dx = \int_{\mathbb{R}}\tilde w_n\Gamma dx\; ,
\end{equation}
where $\{w_n\}_{n\ge 0}$ and $\{\tilde w_n\}_{n\ge0}$ are two families of polynomials in  $\Phi$, $\Gamma$, and their derivatives,  with first elements  $w_0=\Gamma$ and $\tilde w_0 = \Phi$, respectively, so that
\begin{equation}
    Q_0=\int_{\mathbb{R}}\Gamma\Phi dx = N \; .
\end{equation}
For every $n\ge0$, the elements at order $n+1$ are defined recursively in terms of lower order ones
\begin{equation} \label{eq:recursion}
\left\{
\begin{aligned}
    &w_{n+1}=\frac{\mu\sigma^4}{|g|}\left(-\partial_x w_n+\epsilon\,\Phi\,\sum_{k=0}^{n-1}w_kw_{n-k-1}\right)\\
    &\tilde w_{n+1}=\frac{\mu\sigma^4}{|g|}\left(\partial_x \tilde w_n+\epsilon\,\Gamma\,\sum_{k=0}^{n-1}\tilde w_k\tilde w_{n-k-1}\right)
\end{aligned}\right. \; .
\end{equation}
where $\epsilon = {\rm sgn}(g)$.
It is easy to check that the first integrals of motion  this procedure yields are the momentum $P$ (Eq.~\eqref{momentum}) and the Energy $E$ (Eq.~\eqref{energy}) already obtained through Noether theorem
\begin{equation}
Q_1=\frac{1}{2}\int_\mathbb{R} (w_1\Phi+\tilde w_1 \Gamma) dx=\frac{1}{2}\frac{\mu\sigma^4}{|g|}\int_\mathbb{R}( \Gamma\partial_x\Phi-\Phi\partial_x\Gamma) dx \;=\frac{\mu\sigma^4}{|g|} P \; ,
\end{equation}
and
\begin{equation}
Q_2=\frac{1}{2}\int_\mathbb{R} (w_2\Phi+\tilde w_2\Gamma) dx=\frac{\mu^2\sigma^8}{g^2}\int_{\mathbb{R}}dx\left[- \partial_x \Gamma\, \partial_x \Phi+\frac{g}{\mu\sigma^4}\,\left(\Gamma\Phi\right)^2\right] \; = 2 \frac{\mu\sigma^4}{g^2} E\; .
\end{equation}
 Next iterations allow to compute new, less obvious, conserved quantities such as
\begin{equation}
\begin{aligned}
Q_3&=\frac{1}{2}\int_\mathbb{R} (w_3\Phi+\tilde w_3\Gamma) dx\\
&=\frac{1}{2}\frac{\mu^3\sigma^{12}}{|g|^3}\int_\mathbb{R}dx
\left[
\left(\Gamma\partial_{xxx}\Phi-\Phi\partial_{xxx}\Gamma\right)
+3\, \frac{g}{\mu\sigma^4}\, \left( \Gamma^2(\partial_x\Phi)^2-\Phi^2(\partial_x\Gamma)^2\right)
\right]\; ,
\end{aligned}
\end{equation}
This new integral of motion is reminiscent of the generator of the modified KdV operator, which could be expected since mKdV and NLS are both part of the same AKNS hierarchy \cite{ablowitz1981solitons}. We may also compute easily the next integral
\begin{equation}
\begin{aligned}
Q_4&=\frac{1}{2}\int_\mathbb{R} (w_4\Phi+\tilde w_4\Gamma) dx\\
&=\frac{\mu^4\sigma^{16}}{g^4}\int_\mathbb{R}dx\left[\partial_{xx}\Phi\partial_{xx}\Gamma - \frac{g}{\mu\sigma^4}\left( 2\partial_x\Phi^2\partial_x\Gamma^2 + \Phi^2(\partial_x\Gamma)^2+\Gamma^2(\partial_x\Phi)^2\right)+2 \frac{g^2}{\mu^2\sigma^8}(\Phi\Gamma)^3\right] \; .
\end{aligned}
\end{equation}

\section{Integrable systems formalism}
\label{sec:intflism}

\subsection{Nondimensionalization}

As is often the case when dealing with non-linear PDEs, it will prove convenient to parametrize the system of equations \eqref{intNLS} using dimensionless units. 

In analogy with Bose-Einstein condensates, we introduced elsewhere \cite{bonnemain2019universal} the healing length $\nu=\mu\sigma^4/\left|g\right|$ as a typical length scale of the problem. In a similar fashion, we can also define $\tau={2\mu^2\sigma^6}/{g^2}$ as a typical time scale. Denoting both $t'={t}/{\tau}$ and $x'={x}/{\nu}$ we can then write Eqs.~\eqref{intNLS} in a simpler form using dimensionless coordinates
\begin{equation}\label{dimlessNLS}
\left\{
\begin{aligned}
&-\partial_{t'}\Phi=+\partial_{x'x'}\Phi  + 2\,\epsilon\, \nu \,m\,\Phi \\
&+\partial_{t'}\Gamma=+\partial_{x'x'}\Gamma  + 2\,\epsilon\, \nu\, m\,\Gamma
\end{aligned} \right. \; ,
\end{equation}
where the value of $\epsilon=\pm 1$ is given by the sign of the interaction constant  $g$. In this form, the healing length $\nu$ appears as the only relevant parameter.

The above representation (Eqs.~\eqref{dimlessNLS} with all "primes" dropped) will be used for the rest of the paper. It will be especially useful when dealing with the several transformations required by the IST method.

\subsection{Zero curvature representation}

The foundation of the IST method lies in the fact that  Eqs.~\eqref{dimlessNLS} can be seen as compatibility conditions for an auxiliary, overdetermined, linear system. Let $F=(f_1,f_2)$ be a vector function of $(x,t)$ defined by
\begin{equation}\label{auxprob}
\left\{
\begin{aligned}
&\partial_x F = U(x,t,\lambda) F\\
&\partial_t F = V(x,t,\lambda) F\\
\end{aligned}
\right. \; ,
\end{equation}
where $U$ and $V$ are $2\times2$ matrix functions depending on the  space and time variables $x$ and $t$, but also on a \textit{spectral parameter} $\lambda$, the importance of which will be made clear later. By Schwartz's theorem, the two cross derivatives of $F$ have to be equal, which leads to the compatibility condition:
\begin{equation}\label{0curv}
\partial_t U - \partial_x V + [U,V]=0 \; ,
\end{equation}
and this relation has to hold no matter the value taken by $\lambda$. If we assume that
\begin{equation}\label{Upair}
U=\kappa_\epsilon
\begin{pmatrix}
0 & \Phi \\
\Gamma & 0 
\end{pmatrix}
+\begin{pmatrix}
\frac{\lambda}{2} & 0 \\
0 & -\frac{\lambda}{2} 
\end{pmatrix} \; ,
\end{equation}
with $\epsilon=\pm 1$ as in \eqref{dimlessNLS} and, respectively, $\kappa_- = \sqrt{\nu}$  and $\kappa_+ = i\sqrt{\nu}$, and
\begin{equation}\label{Vpair}
V=\kappa_\epsilon
\begin{pmatrix}
\kappa_\epsilon\Phi\Gamma & -\partial_x\Phi \\
\partial_x\Gamma & -\kappa_\epsilon\Phi\Gamma
\end{pmatrix} -\lambda U \; ,
\end{equation}
then, under the constraints Eqs.~\eqref{eq:PhiGammaToZero}, Eqs.~\eqref{dimlessNLS} are equivalent to the compatibility condition \eqref{0curv}. More general boundary conditions can be accounted for by modifying $V$ (see for instance the chapter \textit{The case of finite density} in \cite{faddeev2007hamiltonian}). What this representation brings is a fairly natural geometric interpretation. The matrices $U$ and $V$ can be seen as the $x$ and $t$ components of a connection (or gauge field) in the vector bundle $\mathbb{R}^2\times\mathbb{R}^{+2}$, while the left hand side of compatibility condition \eqref{0curv} can be seen as the curvature (or strength field) of this connection according to the Ambrose-Singer theorem \cite{ambrose1953theorem}. Hence the name \textit{zero-curvature representation}. In the field of classical integrable systems this is known as Lax connection and the compatibility equation \eqref{0curv} is equivalent to Lax equation in the limit of an infinite number of degrees of freedom \cite{babelon2003introduction}.

\subsection{Parallel transport}

A reasonable progression, once we interpret $(U,V)$ as a connection, is to consider the parallel transport it induces. Let $\gamma$ be a curve in $\mathbb{R}^2$  and $\gamma_1...\gamma_N$ a partition into $N$ adjacent segments. We define the parallel transport along $\gamma$ as
\begin{equation}\label{parTpartition}
\Omega_\gamma = \lim_{N\rightarrow\infty}\left[\mathcal{P}\prod_{n=1}^{N}\left(\mathbb{1}+\int_{\gamma_n}(Udx+Vdt)\right)\right] \; ,
\end{equation} 
where $\mathcal{P}$ denotes path ordering and $\mathbb{1}$ the $2\times2$ identity matrix. A more compact notation for this expression would be
\begin{equation}\label{parT}
\Omega_\gamma=\mathcal{P}\exp\int_{\gamma}(Udx+Vdt) \; .
\end{equation}
This last expression is also particularly convenient for two reasons. The first one is that it makes it clear that if $\gamma$ represents a path from some point  $(x,s)$  to some other point $(y,t)$,  and given the initial data $F(x,s)$, the solution of equations \eqref{auxprob}  can be written as a covariantly constant vector field
\begin{equation}\label{covectfield}
F(y,t)=\Omega_\gamma F(x,s) \; .
\end{equation}
The second, and maybe more important, reason is that it  shows clearly that parallel transport over any closed curve $\gamma_0$ (holonomy of the connection) is trivial by way of the non-Abelian Stokes theorem (cf appendix \ref{app:nabstokes}). Indeed, thanks to the vanishing of the curvature,
\begin{equation}\label{holonomy}
\Omega_{\gamma_0}=\mathbb{1} \; .
\end{equation}
This property is akin to that of Lagrangian manifolds in Hamiltonian mechanics and will prove to play a fundamental role in the computations of integrals of motion.

\subsection{Monodromy matrices}

\begin{figure}[h]
	\includegraphics[width=0.97\textwidth]{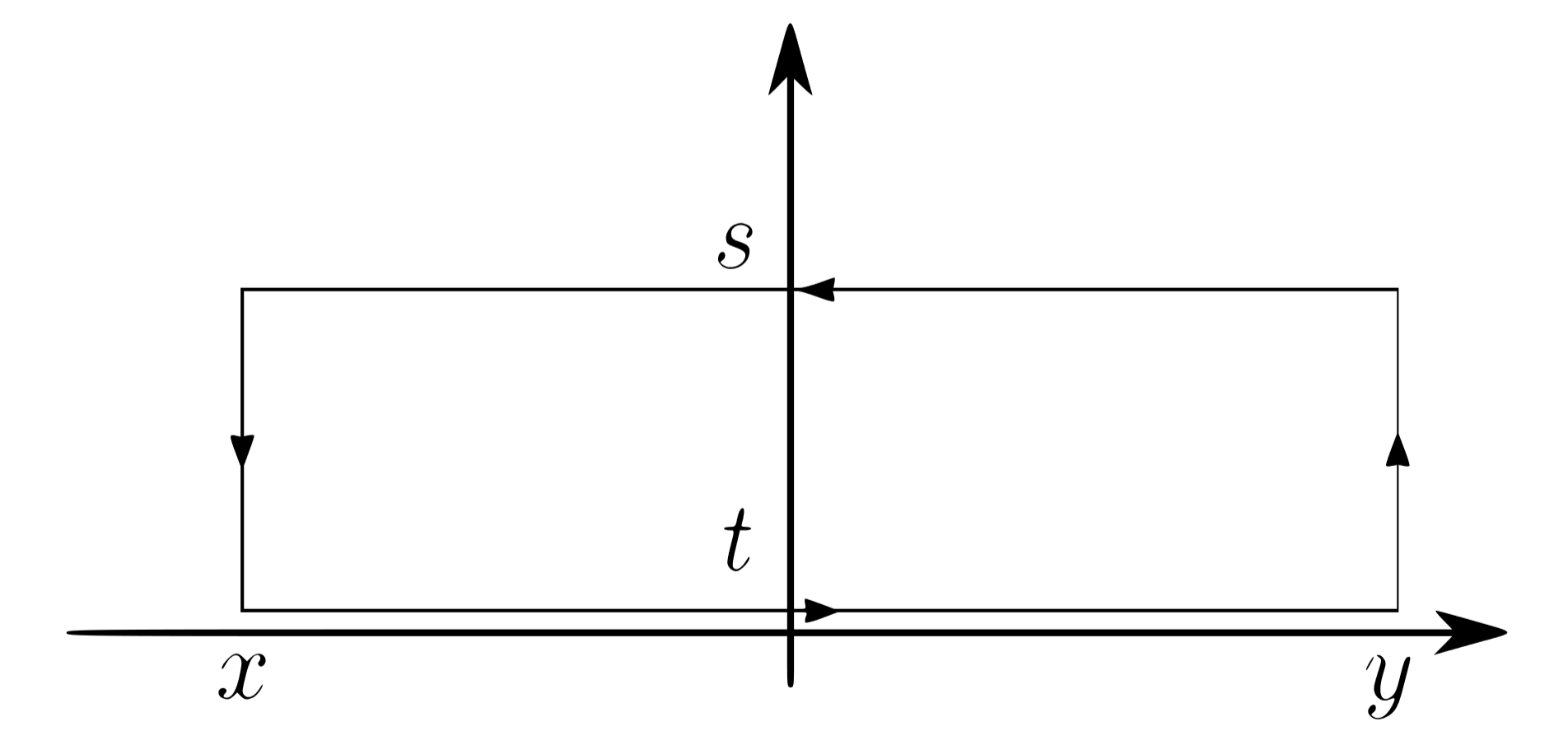}
	\caption{A rectangular loop $\gamma_R$ in the two dimensional space-time. Vanishing of the curvature imposes that parallel transport along $\gamma_R$ is trivial.} 
	\label{fig:loop}
\end{figure}

The main characteristics of the problem are the two \textit{monodromy matrices} defined respectively as the propagators in the  space and time directions
\begin{equation}\label{monodef}
\left\{
\begin{aligned}
&T(x,y,\lambda;\tau)=\mathcal{P}\exp\int_{x}^y U(z,\tau,\lambda) dz \\
&S(s,t,\lambda;z) = \mathcal{P}\exp\int_{s}^{t} V(z,\tau,\lambda) d\tau
\end{aligned}
\right. \; .
\end{equation}
These "global" objects will turn out to be easier to manipulate than their local counterparts $U$ and $V$, notably thanks to the non-Abelian Stokes theorem. To illustrate this, we shall consider a closed rectangular loop $\gamma_R$ as represented Figure \eqref{fig:loop}.
Because of its geometry,  parallel transport along $\gamma_R$ can be readily expressed in terms of the monodromy matrices
\begin{equation}\label{monoloop}
\begin{aligned}
\Omega_{\gamma_R}&=S(t,s,\lambda;x)T(y,x,\lambda;t)S(s,t,\lambda;y)T(x,y,\lambda;s)=\mathbb{1}
\end{aligned}\; .
\end{equation}
 As a particular case, we get  the following inversion property
\begin{equation}\label{invmono}
\left\{
\begin{aligned}
&T(y,x,\lambda;\tau)= T^{-1}(x,y,\lambda;\tau)\\
&S(t,s,\lambda;z)=S^{-1}(s,t,\lambda;z)
\end{aligned}
\right. \; ,
\end{equation}
which in turn can be used to write Eq.~\eqref{monoloop} as
\begin{equation}\label{Tgauge}
T(x,y,\lambda;t)=S(s,t,\lambda;y)\;T(x,y,\lambda;s)\;S^{-1}(s,t,\lambda;x) 
\end{equation}
 This expression is particularly useful whenever the two points $x$ and $y$  are such that for all values of $\lambda$ and all times  $\tau \in [s,t]$, one has:
\begin{equation}\label{VVV}
V(x,\tau,\lambda)=V(y,\tau,\lambda) \; ,
\end{equation}
 In such a case, $S(s,t,\lambda;x) = S(s,t,\lambda;y)$ and  equation \eqref{Tgauge} implies that the time evolution of the monodromy matrix $T(x,y,\lambda)$  just amounts  to a gauge transformation.  In particular, we get that the trace of the monodromy matrix  $\mathrm{Tr} \left[T(x,y,\lambda)\right]$ is  constant in time for all $\lambda$
\begin{equation}\label{wilsline}
\begin{aligned}
\mathrm{Tr}\left[T(x,y,\lambda;t)\right]&=\mathrm{Tr}\left[S(s,t,\lambda;y)\;T(x,y,\lambda;s)\;S^{-1}(s,t,\lambda;x)\right]\\
&=\mathrm{Tr}\left[T(x,y,\lambda;s)\right] \; 
\end{aligned}
\end{equation}
 In what follows, we will use this property to build a generating function for the constants of motion. 

\section{First integrals of the motion}
\label{sec:1stint}

\subsection{Computing conserved quantities}\label{sec:abelian}

 In order to compute all the conserved quantities one first needs to write the monodromy matrix as a Poincaré expansion in $\lambda$ \cite{faddeev2007hamiltonian}. 

We  first introduce the monodromy matrix $E(y-x,\lambda)$ associated with Eqs.~\eqref{dimlessNLS} for the trivial constant solution $\Phi(x,t)=0$, $\Gamma(x,t)=0$
\begin{equation}\label{E}
\begin{aligned}
E(y-x,\lambda)&=\lim_{\Phi\to 0}\lim_{\Gamma\to 0}\left[\mathcal{P}\exp\int_{x}^yU(z,t,\lambda)dz\right] \\
&=\exp\left[\frac{\lambda}{2}(y-x)\sigma_3\right] \; ,
\end{aligned}
\end{equation}
where 
\begin{equation}\label{Pauli3}
\sigma_3=
\begin{pmatrix}
1 & 0 \\
0 & -1
\end{pmatrix} \; .
\end{equation}
The monodromy matrix can then be written as an expansion in inverse powers of $\lambda$
\begin{equation}\label{monexp}
\begin{aligned}
T(x,y,\lambda;t)&=E(y-x,\lambda)+\sum_{n=0}^{\infty}\frac{T_n(x,y;t)E(y-x,\lambda)}{\lambda^{n+1}}\\
&+\sum_{n=0}^{\infty}\frac{\tilde{T}_n(x,y;t)E(x-y,\lambda)}{\lambda^{n+1}} \; . 
\end{aligned}
\end{equation}
To each order in this expansion  we will show that there is an associated conserved quantity. 

To achieve this we look for an expression for the monodromy matrix in the form
\begin{equation}\label{abelianization}
T(x,y,\lambda;t)=\left(\mathbb{1}+W(y,\lambda;t)\right)\exp Z(x,y,\lambda;t)\left(\mathbb{1} + W(x,\lambda;t)\right)^{-1} \; ,
\end{equation}
where $W$ and $Z$ are respectively an off-diagonal and a diagonal matrix, with the following Poincaré expansions \cite{faddeev2007hamiltonian}
\begin{equation}\label{Wexp}
W(y,\lambda;t)=\sum_{n=0}^{\infty}\frac{W_n(y;t)}{\lambda^{n+1}} \; ,
\end{equation} 
and
\begin{equation}\label{Zexp}
Z(x,y,\lambda;t)=E(y-x,\lambda)+\sum_{n=0}^{\infty}\frac{Z_n(x,y;t)}{\lambda^{n+1}}\; .
\end{equation}
 Starting from the first order differential equation fulfilled by the monodromy matrix $T$:
\begin{equation}\label{Tdiffeq}
\partial_y T(x,y,\lambda;t) = U(y,t,\lambda) T(x,y,\lambda;t) \; ,
\end{equation}
with initial condition
\begin{equation}\label{Tinit}
T(x,y,\lambda;t)|_{x=y}=\mathbb{1} \; ,
\end{equation}
we solve recursively this equation using the representation \eqref{abelianization}, proving in turn its validity.

Inserting the expression \eqref{abelianization} in equation \eqref{Tdiffeq}, and separating diagonal and off-diagonal parts, one obtains
\begin{equation}\label{diagoffeq}
\left\{
\begin{aligned}
&\partial_yZ(x,y,\lambda;t)=\frac{\lambda}{2}\sigma_3+ U_0 (y,t)W(y,\lambda;t)\\
&\partial_yW(y,\lambda;t)+W(y,\lambda;t)\partial_y Z(x,y\lambda;t)= U_0(y,t)+\frac{\lambda}{2}\sigma_3W(y,\lambda;t)
\end{aligned}
\right. \; ,
\end{equation}
where we have used the shorthand notation $U_0 (x,y) \equiv U(y,t,0)$.  Eliminating $Z(x,y,\lambda;t)$ between the two equations  \eqref{diagoffeq}  on gets that $W(y,\lambda;t)$ is solution of a  Riccati equation
\begin{equation}\label{Wdiffeq}
\partial_y W-\lambda\sigma_3W+W U_0 W- U_0 =0 \; .
\end{equation}
Using expansion \eqref{Wexp} one gets a solution of \eqref{Wdiffeq} as a recursion relation:
\begin{equation}\label{Wrec}
W_{n+1}=\sigma_3\left[\partial_y W_n+\sum_{k=0}^{n-1}W_k U_0 W_{n-k}\right] \; ,
\end{equation}
with 
\begin{equation}\label{Winit}
W_0=-\sigma_3  U_0 \; .
\end{equation}
More explicitly, if we write
\begin{equation}\label{wexp}
W=\kappa_\epsilon
\begin{pmatrix}
0 & \displaystyle{-\sum_{n=0}^{\infty}\frac{1}{\lambda^{n+1}}\,\tilde w_n }\\
\displaystyle{\sum_{n=0}^{\infty}\frac{1}{\lambda^{n+1}}\,w_n} & 0
\end{pmatrix} \; ,
\end{equation}
the recursion relation \eqref{Wrec} becomes
\begin{equation}\label{wrec}
\left\{
\begin{aligned}
&w_{n+1}=-\partial_y w_n +\epsilon\,\nu\, \Phi\sum_{k=0}^{n-1}w_kw_{n-k-1}\\
&\tilde w_{n+1}=\partial_y\tilde w_n +\epsilon\,\nu\, \Gamma\sum_{k=0}^{n-1}\tilde w_k\tilde w_{n-k-1}\\
\end{aligned}
\right. \; ,
\end{equation}
with
\begin{equation}\label{winit}
\left\{
\begin{aligned}
&w_{0}=\Gamma\\
&\tilde w_{0}=\Phi\\
\end{aligned}
\right. \; .
\end{equation}
Now, the first equation in \eqref{diagoffeq} can be readily integrated as
\begin{equation}\label{Zeq}
Z(x,y,\lambda;t)=\frac{\lambda(y-x)}{2}\,\sigma_3+\int_{x}^{y}U_0(z,t)W(z,\lambda;t)dz \; .
\end{equation}
 Using expansion \eqref{wrec} this last expression becomes
\begin{equation}\label{Zmatrix}
\begin{aligned}
Z(x,y,\lambda;t)&=\frac{\lambda(y-x)}{2}
\begin{pmatrix}
1 & 0 \\
0 & -1
\end{pmatrix}\\
&+ \kappa_\epsilon\,
\begin{pmatrix}
\displaystyle{\sum_{n=0}^{\infty}\frac{1}{\lambda^{n+1}}\int_{x}^{y}w_n\Phi dz} & 0 \\
0 & \displaystyle{-\sum_{n=0}^{\infty}\frac{1}{\lambda^{n+1}}\int_{x}^{y}\tilde w_n\Gamma dz}
\end{pmatrix}
\end{aligned} \; 
\end{equation}
By way of equations \eqref{wrec} and \eqref{winit}, it is easy to check that for all $n$
\begin{equation}\label{wtildew}
\int_{x}^{y}w_n(z,t)\Phi(z,t) dz=\int_{x}^{y}\tilde w_n(z,t)\Gamma(z,t) dz \; 
\end{equation}
Thus we have
\begin{equation}\label{Zmatrix2}
Z(x,y,\lambda;t)=\left[\frac{1}{2}\lambda(y-x)
+ \kappa_\epsilon^2\,
\sum_{n=0}^{\infty}\frac{1}{\lambda^{n+1}}\int_{x}^{y}w_n(z,t)\,\Phi(z,t)\, dz
\right] \sigma_3
\end{equation}
and we obtain from equation \eqref{abelianization}
\begin{equation}\label{genfun}
\begin{aligned}
\mathrm{Tr}\left[T\right]&=\mathrm{Tr}\left[\exp Z\right]\\
&=2\mathrm{ch}\left[\lambda(y-x) -\epsilon \nu \sum_{n=0}^{\infty} \frac{1}{\lambda^{n+1}} \int_{x}^{y}w_n(z,t)\Phi(z,t) dz\right]
\end{aligned}\; ,
\end{equation}
meaning that for all $n$ the integral \eqref{wtildew} has to be constant in time. Integrals of motion are obtained by computing the monodromy matrix over the whole space, letting the interval $]x,y[\rightarrow \mathbb{R}$.  In this limit, the first term inside the right hand side of  Eq.~\eqref{genfun} diverges but this can be dealt with using a standard renormalisation procedure (cf Appendix \ref{app:dynmon}). 
Up to a $\nu^{-1}$ factor due to rescaling of lengths \eqref{dimlessNLS}, one finds the same conserved quantities obtained through \eqref{eq:Qn}. Indeed the first three quantities read 
\begin{equation}\label{intmotion}
\begin{aligned}
&Q_0=\int_\mathbb{R} w_0\Phi dx=\int_\mathbb{R} \Gamma\Phi dx
 \,=\, \frac{1}{\nu}\, N \\
&Q_1=\int_\mathbb{R} w_1\Phi dx=\frac{1}{2}\int_\mathbb{R}( \Gamma\partial_x\Phi-\Phi\partial_x\Gamma) dx \,=\, \, P\\
&Q_2=\int_\mathbb{R} w_2\Phi dx=\int_\mathbb{R}(  -\partial_x\Phi\partial_x\Gamma
 +\epsilon\, \nu\, \Phi^2\,\Gamma^2) dx  \,=\, \frac{2}{|g|} \, E
\end{aligned} \qquad ,
\end{equation}
and one can check that these are indeed equivalent to the conserved quantities $N$, $P$ and $E$ introduced in section \ref{sec:noether}. Higher order terms correspond to more abstract quantities we will not discuss but are still, by construction, invariant.

\subsection{Poisson commutativity of the first integrals of motion} \label{sec:HamFor}

For the MFG equations \eqref{iqMFG} to be completely integrable in the Liouville sense, the (infinite number of) conserved quantities generated in section \ref{sec:abelian} need to be in involution. Here, we introduce a Poisson structure in the context of integrable MFGs and use it to show the Poisson commutativity of the aforementioned conserved quantities. Here again this amounts to adapting a standard procedure \cite{babelon2003introduction}, presented in the present context for the sake of completeness.

\subsubsection{Generalisation of Poisson brackets to infinite dimensional systems}

For N-dimensional Hamiltonian systems, given two functions $f(p_i,q_i,t)$ and $g(p_i,q_i,t)$ of Darboux coordinates $(p_i,q_i)$ on the phase space, \textit{Poisson brackets} take the form
\begin{equation}\label{finitePB}
\{f,g\}=\sum_{i=1}^{N}\left(\frac{\partial f}{\partial q_i}\frac{\partial g}{\partial p_i}-\frac{\partial f}{\partial p_i}\frac{\partial g}{\partial q_i}\right) \; .
\end{equation} 
However, MFG equations \eqref{iqMFG} constitute an infinite-dimensional system and definition \eqref{finitePB} needs to be extended. In this case the phase space $\mathcal M$ is an infinite-dimensional real space with positive coordinates defined by pairs of functions $\Phi(x,t)$ and $\Gamma(x,t)$ \footnote{By analogy with finite-dimensional coordinates, x may be thought of a coordinate label.}. On this phase space, the algebra of observables is made up of smooth, real, analytic functionals, on which one can define a \textit{Poisson structure} by the following bracket
\begin{equation}\label{infPB}
\{F,G\}=\int_{\mathbb R}\left(\frac{\delta F}{\delta \Gamma} \frac{\delta G}{\delta \Phi}-\frac{\delta F}{\delta \Phi}\frac{\delta G}{\delta \Gamma}\right)dx \; ,
\end{equation} 
which possesses the standard properties of Poisson brackets: it is skew-symmetric and satisfies Jacobi identity. The coordinates $\Phi$ and $\Gamma$ may themselves be considered functionals on $\mathcal M$ such that
\begin{equation}\label{coobrack}
\begin{aligned}
&\{ \Gamma(x,t),\Phi(y,t) \}
=\delta(x-y) \\
&\{\Gamma(x,t),\Gamma(y,t)\}=\{\Phi(x,t),\Phi(y,t)\}=0
\end{aligned}\; .
\end{equation}
These formulae directly yield that, for any observable $F$
\begin{equation}\label{obsvar}
\{\Gamma,F\}=\frac{\delta F}{\delta \Phi} \quad \text{and} \quad \{\Phi,F\}=-\frac{\delta F}{\delta \Gamma} \; ,
\end{equation}
and in particular, if one takes as  observable  the conserved quantity $Q_2$ which is proportional to the energy $E$ (Eq.~\eqref{energy}), one gets the equations of motion in Hamiltonian form:
\begin{equation}\label{hamform}
\left\{
\begin{aligned}
&\partial_t \Gamma= \{\Gamma,Q_2\}=\frac{\delta Q_2}{\delta \Phi}\\
&\partial_t \Phi = \{\Phi,Q_2\}= -\frac{\delta Q_2}{\delta \Gamma}
\end{aligned}
\right. \; ,
\end{equation}
which are equivalent to MFG equations \eqref{dimlessNLS}. The Poisson structure defined by the non-degenerate bracket \eqref{infPB} highlights the symplectic nature of the phase space $\mathcal M$ and each of the Poisson commuting integrals of motion correspond to a sheet of the regular foliation of this phase space. This provides yet another, Hamiltonian, representation of MFG problems.

\subsubsection{Classical r-matrix}

The simplest way to check that all the invariant observables generated section \ref{sec:abelian} are in involution (and prove that the system is completely integrable in the Liouville sense) is probably to verify that the Poisson bracket of the trace of the monodromy matrix with itself vanishes
\begin{equation}\label{toprove}
\{\mathrm{Tr}\left[T\right],\mathrm{Tr}\left[T\right]\}=0 \; ,
\end{equation}
as $\mathrm{Tr}\left[T\right]$ can serve as generating functions for the constant of motion. In this section we will introduce a powerful tool that will help us with these computations: the \textit{classical r-matrix}.

To that end, let us define a \textit{tensorial Poisson bracket} for any $2\times2$ matrix functionals $A$ and $B$ \footnote{This can naturally be generalized to $n\times n$ matrices, but we restrict the discussion to matrices of the size of $T$.}
\begin{equation}\label{tensPB}
\{A\otimes B\} = \int_{\mathbb R} \left(
\frac{\delta A}{\delta \Gamma}\otimes \frac{\delta B}{\delta \Phi}-\frac{\delta A}{\delta \Phi}\otimes \frac{\delta B}{\delta \Gamma}
\right) dx \; ,
\end{equation}
such that
\begin{equation}\label{tensPBel}
\{A\otimes B\}_{(j,k),(m,n)} = \{A_{j,m},B_{k,n}\} \; .
\end{equation}

Hence, using relations \eqref{coobrack}, one can compute the bracket of $U$, $x$-component of the Lax connection, with itself
\begin{equation}\label{UPB}
\{U(x,\lambda)\otimes U(y,\mu)\}=\nu(\sigma_-\otimes\sigma_+-\sigma_+\otimes\sigma_-)\delta(x-y) \; ,
\end{equation}
where
\begin{equation}\label{Pauli+}
\sigma_-=
\begin{pmatrix}
0 & 0 \\
1 & 0
\end{pmatrix} \quad \text{and} \quad
\sigma_+=
\begin{pmatrix}
0 & 1 \\
0 & 0
\end{pmatrix}\; .
\end{equation}
Equation \eqref{UPB} can also be written as a commutator
\begin{equation}\label{ultraloc}
\{U(x,\lambda)\otimes U(y,\mu)\}=\bigl[r(\lambda-\mu)\,,\,U(x,\lambda)\otimes\mathbb{I}+
\mathbb{I}\otimes U(y,\mu)
\bigr]\,\delta(x-y) \; ,
\end{equation}
involving the classical r-matrix which here takes the form
\begin{equation}\label{rmat}
r(\lambda)=-\frac{\nu}{\lambda}\begin{pmatrix}
1 & 0 & 0 & 0 \\
0 & 0 & 1 & 0 \\
0 & 1 & 0 & 0 \\
0 & 0 & 0 & 1 
\end{pmatrix} \; .
\end{equation}
The point of this formulation is to express tensorial Poisson brackets, which may be difficult to compute for the monodromy matrix $T$, as simple commutators. The existence of the $r$-matrix and formulation \eqref{ultraloc} underlies integrability and has a natural Lie-algebraic interpretation. As such, this relation takes the name of \textit{fundamental Poisson bracket}.

\subsubsection{Sklyanin fundamental relation}

To compute the Poisson bracket of $T$ with itself one can evaluate an integral version of the fundamental Poisson bracket \eqref{ultraloc}
\begin{equation}\label{Tbrack}
\begin{aligned}
\{T_{ab}&(x,y,\lambda),T_{cd}(x,y,\mu)\}=\\
&\int_{x}^{y}\frac{\delta T_{ab}(x,y,\lambda)}{\delta U_{jk}(z,\lambda)}\{U_{jk}(z,\lambda),U_{lm}(z',\mu)\}\frac{\delta T_{cd}(x,y,\mu)}{\delta U_{lm}(z',\mu)}dzdz' \; .
\end{aligned}
\end{equation}
By varying the differential equation \eqref{Tdiffeq}, that serves as definition of the monodromy matrix $T$
\begin{equation}\label{varTdiffeq}
\partial_x \delta T(x,y,\lambda)=\delta U(x,\lambda)T(x,y,\lambda)+U(x,\lambda)\delta T(x,y,\lambda) \; ,
\end{equation}
the solution of which is
\begin{equation}\label{deltaT}
\delta T(x,y,\lambda)=\int_{y}^{x}T(x,z)\delta U(z)T(z,y)dz \; ,
\end{equation}
it follows that
\begin{equation}\label{varderT}
\frac{\delta T_{ab}(x,y,\lambda)}{\delta U_{jk}(z,\lambda)}=T_{aj}(x,z,\lambda)T_{kb}(z,y,\lambda) \; .
\end{equation}
Inserting this last expression in equation \eqref{Tbrack} one eventually gets
\begin{equation}\label{RTTint}
\begin{aligned}
\{T(x,y,\lambda)\otimes T(x,y,\mu)\}= &\int_{y}^{x}\left(T(x,z,\lambda)\otimes T(x,z,\mu)\right)\\
&[r(\lambda-\mu),U(z,\lambda)\otimes\mathbb{I}+\mathbb{I}\otimes U(z,\mu)]\\
&\left(T(z,y,\lambda)\otimes T(z,y,\mu)\right) dz
\end{aligned} \; ,
\end{equation}
that simplifies, noticing the integrand is a total derivative with respect to $z$,
\begin{equation}\label{RTT}
\{T(x,y,\lambda)\otimes T(x,y,\mu)\}=-[r(\lambda-\mu),T(x,y,\lambda)\otimes T(x,y,\mu)] \; .
\end{equation}
This formulation is sometimes called \textit{RTT Poisson structure} or \textit{Sklyanin fundamental relation}.

\subsubsection{Involution of the first integrals of motion}

From Sklyanin fundamental relation (now that integrals depending on unknown fields $\Phi$ and $\Gamma$ that constitute Poisson brackets are dealt with implicitly) it is easy to show the constants of motion are in involution. In particular, recalling that, for any pair of matrices $(A,B)$
\begin{equation}\label{tensrel}
\mathrm{Tr}(A\otimes B)=\mathrm{Tr}[A]\mathrm{Tr}[B] \; ,
\end{equation}
yields from equation \eqref{RTT}, since the trace of a commutator is zero,
\begin{equation}\label{inv}
\{\mathrm{Tr}[T(x,y,\lambda)],\mathrm{Tr}[T(x,y,\mu)]\}=0 \; ,
\end{equation}
proving the involution of the first integrals of motion.

The fact that all the conserved quantities generated by equation \eqref{genfun} are in involution implies in turn the existence of an infinite  set of systems of dynamical equations of Hamilton-Poisson type
\begin{equation}\label{hier}
\left\{
\begin{aligned}
&\partial_t \Gamma= \{\Gamma,Q_k\} \\
&\partial_t \Phi= \{\Phi,Q_k\}
\end{aligned}
\right. \; ,
\end{equation}
that all have the same integrals of motion $\{Q_n\}_{n\in\mathbb{N}}$. This directly leads to a whole family of new integrable Forward-Backward equations of MFG-type, which may prove interesting in their own right.

\section{Conclusion}

In this paper, we have used a formal connection between Quadratic Mean Field Games (MFG) and the Non-Linear Schrödinger (NLS) equations to analyse  an instance of the former, Eq \eqref{iqMFG}, as an integrable system.  In  the present approach, we have been able to determine an infinite set of constants of  motion  through recursion relations, Eqs \eqref{eq:Qn}-\eqref{eq:recursion}, and prove that they are in mutual involution in Liouville sense.  The  first three terms of this hierarchy of first integrals can be interpreted as the MFG analogs of the total mass, momentum and energy, respectively.

In the known NLS case, it is possible to go beyond the identification of these first integrals, and obtain an essentially complete analytical solution for the corresponding integrable limit through the Inverse Scattering Transform (IST) approach, a powerful tool that can be seen as a sophisticated Fourier transform  for non-linear equations. At present, it is not fully clear whether this program can be fully transposed to the  Mean Field Games context, but it would provide further understanding of  Eqs.~\eqref{intNLS}, and give insights on their forward-backward structure. Such an implementation as a method for solving MFG equation can be divided in three major steps, just like Fourier transform for translation invariant systems, which are summarized in the following commutative diagram:
\begin{equation*}
	\xymatrix@C=4cm@R=2cm{
		\text{$[\Phi(T,x), \Gamma(0,x)]$}\ar[r]^-{\text{\large IST}}\ar@{.>}[d]_{\vphantom{\text{(}}\text{dynamics}}^{\text{(complicated)}} 
		& \text{Scattering data}\ar[d]_{\vphantom{\text{(}}\text{dynamics}}^{\text{(simple)}}\\
		\text{$[\Phi(t,x),\Gamma(t,x)]$} 
		& \text{Evolved scattering data}\ar[l]_-{\text{\large IST$^{-1}$}}} \; .
\end{equation*}
First, one needs to relate the fields $\Phi$ and $\Gamma$ to their associated scattering data (the integrals of motion being part of these data), all of which can be extracted from the monodromy matrix $T$. In terms of Hamiltonian mechanics, the IST essentially defines a canonical transformation to action-angle variables. In a second step, one has to compute the time evolution of these scattering data, which is significantly simpler than the original equations \eqref{intNLS} as the dynamics of $T$ is actually linear \cite{bonnemainthesis} (cf. Appendix \ref{app:dynmon}). The last step is also the most arduous one. It consists in reconstructing the fields from the evolved scattering data, and this usually amounts to solving some instance of the Riemann-Hilbert problem or Gelfand-Levitan-Marchenko integral equation \cite{faddeev2007hamiltonian,babelon2003introduction,ablowitz1974inverse}.

This paper only provides a modest contribution towards the ambitious goal of constructing the inverse scattering transform for integrable mean field games: it introduces the formalism of integrable systems in the field of Mean Field Games and and highlights an infinite set of conserved quantities. However, showing the existence of a zero-curvature representation is already sufficient to pursue this program since the vanishing curvature of the Lax connection underlies a Poisson structure. Eqs.~\eqref{intNLS} constitute an infinite-dimensional Hamiltonian system for which there exist action-angle coordinates and therefore a transform similar to IST should exist in the context of MFG. However, new technical difficulties arise in this context, the most important are the two following ones:
\begin{enumerate}
    \item In standard NLS equation, the two fields $\Psi$ and $\bar \Psi$, corresponding here to $\Phi$ and $\Gamma$, are complex conjugate. It naturally induces a symmetry between the elements of the monodromy matrix, which is of great help in obtaining scattering data, as well as constructing inverse transformation. Such a symmetry does not exist in the context of MFG.
    \item It is not yet clear how the issue of the forward-backward structure of Eqs~\eqref{intNLS} would affect the IST but we can see two avenues to attempt solving that problem. The first one would be to add a self-consistency equation, on top of the IST, but this may prove to be highly non-trivial to solve. Another solution would be to try and study the monodromy matrix in time along with $T$ and use the notion of duality of Lax pairs as discussed in \cite{avan2017origin}. Both approaches seem equally reasonable to follow but appear to require a better understanding of the forward-backward structure.
\end{enumerate}
We plan to address these items in subsequent works.

\appendix

\section{Non-Abelian Stokes theorem}
\label{app:nabstokes}

The aim of this appendix is to introduce the non-Abelian Stokes theorem in the context of the IST. Here we stick to a rather concise discussion, and more details can be found in \cite{broda2001non}.

\subsection{Stokes theorem}

We start by briefly recalling the traditional, Abelian, Stokes theorem. Let $N$ be a $d$-dimensional manifold, $\partial N$ its $(d-1)$-dimensional boundary and $\omega$ a $(d-1)$-form with differential $d\omega$. Then this theorem states that
\begin{equation}\label{StokesThm}
\int_N d\omega=\int_{\partial N}\omega \; ,
\end{equation}
converting an integral over a closed surface into a volume integral. 

\subsection{Generalization to non Abelian forms}

To generalize the previous result, one can introduce the covariant derivative
\begin{equation}\label{covdev}
D_i = \partial_i - A_i \; ,
\end{equation}
where $A_i$ is the $i$ component of a connection. Then, the non Abelian version of the relation \eqref{StokesThm} naturally reads
\begin{equation}\label{nabStokesThm}
\mathcal{P}\exp\oint A=\mathcal{P}\exp\int DA \; ,
\end{equation}
where $\mathcal{P}$ denotes the path ordering. Now, let us recall the compatibility condition \eqref{0curv} of the auxiliary problem \eqref{auxprob}
\begin{equation}
\partial_t U - \partial_x V + [U,V]=0 \; .
\end{equation}
As mentioned earlier, $U$ and $V$ can be interpreted as a connection (or gauge potential), used to define the parallel transport $\Omega$ through equation \eqref{parT}. We can make this more explicit by noticing that the compatibility condition \eqref{0curv} can be rewritten in a very compact way
\begin{equation}\label{0compact}
[D_0,D_1]=0 \; ,
\end{equation}
with
\begin{equation}\label{covcomp}
\left\{
\begin{aligned}
&\partial_0-A_0=\partial_x-U\\
&\partial_1-A_1=\partial_t-V
\end{aligned}
\right. \; ,
\end{equation}
which is equivalent to saying that the differential form $A=A_idx^i$ has a vanishing covariant derivative
\begin{equation}\label{vancoder}
DA=D_jA_idx^i\wedge dx^j=0 \; .
\end{equation}
By way of the non-Abelian Stokes theorem, this means that
\begin{equation}\label{parT1}
\mathcal{P}\exp\oint A=\mathbb{1} \; ,
\end{equation}
hence the name \textit{zero-curvature condition}. 

\section{Dynamics of the monodromy matrix}\label{app:dynmon}

In order to study the time evolution of the monodromy matrix we shall take the time derivative of equation \eqref{Tdiffeq}
\begin{equation}\label{Tdiffxt}
\partial_{t,y}T=\partial_t UT-U\partial_t T \; .
\end{equation}
Using the compatibility conditions \eqref{0curv} to express $\partial_t U$ in terms of $V$ yields
\begin{equation}\label{Tdiffxt0curv}
\partial_y(\partial_t T-VT)=U(\partial_t T-VT)\; ,
\end{equation}
from which we can infer 
\begin{equation}\label{Tdifft}
\partial_t T(x,y,\lambda;t)=V(y,t,\lambda)T(x,y,\lambda;t)+T(x,y,\lambda;t)V(x,t,\lambda) \; ,
\end{equation}
by making use of the initial condition \eqref{Tinit}. 

Let us now consider the monodromy over the whole domain. Introducing the reduced monodromy matrix
\begin{equation}\label{redmon}
\begin{aligned}
\mathcal{T}(\lambda,t)&\equiv\lim_{x\rightarrow+\infty}\lim_{y\rightarrow-\infty}E(-x,\lambda)T(x,y,\lambda;t)E(y,\lambda) \equiv
\begin{pmatrix}
a(\lambda,t) & b(\lambda,t) \\
c(\lambda,t) & d(\lambda,t)
\end{pmatrix}
\end{aligned} \; ,
\end{equation}
the asymptotic behaviour
\begin{equation}\label{asyV}
\lim_{x\rightarrow\pm\infty}V(x,t,\lambda)E(x)=\frac{\lambda^2}{2}\sigma_3E(x)
\end{equation}
leads to the remarkably simple dynamics
\begin{equation}\label{redmonT}
\partial_t \mathcal{T}(\lambda,t)=\frac{\lambda^2}{2}[\sigma_3,\mathcal{T}] \; .
\end{equation}
A particularly interesting aspect of this equation is that the explicit dependence on $\Phi$ and $\Gamma$ has completely disappeared, making for a trivial resolution. In terms of the coefficients of $\mathcal{T}$ this means that the diagonal coefficients $a$ and $d$  are constant in time
\begin{equation}\label{transcoeffdiag}
\begin{aligned}
&a(\lambda,t)=a(\lambda,0) \\
&d(\lambda,t)=d(\lambda,0)
\end{aligned} \; ,
\end{equation}
which was to be expected based on equation \eqref{genfun}. Moreover, the off-diagonal coefficients $b$ and $c$ can be expressed as
\begin{equation}\label{transcoeffoffdiag}
\begin{aligned}
&b(\lambda,t)=b(\lambda,0)e^{\lambda^2t} \\
&c(\lambda,t)=c(\lambda,0)e^{-\lambda^2t}
\end{aligned} \; .
\end{equation}
This simplification comes from the fact that the IST can be interpreted, from the Hamiltonian standpoint, as a transformation to action-angle variables. The trace of $\mathcal{T}$  is a generating functional for the conserved quantities (and hence for the action variables), while the off-diagonal elements play the role of angle variables.

\bibliography{Biblio,Biblio-2019-01-03,Bibliography}
\end{document}